\documentclass[9pt,twocolumn,twoside]{pnas-new}

\usepackage{bm}  
\usepackage{bigstrut}  
\usepackage{makecell}
\usepackage[normalem]{ulem}

\newcommand{\eref}[1]{\eqref{#1}}
\newcommand{\esref}[1]{\eqref{#1}}
\newcommand{\seref}[1]{\eqref{#1}}

\newcommand{\fref}[1]{Figure~\ref{#1}}

\newcommand{\tref}[1]{Table~\ref{#1}}

\templatetype{pnasresearcharticle} 

\title{A minimal-length approach unifies rigidity in under-constrained materials}

\author[a,1]{Matthias Merkel}
\author[b]{Karsten Baumgarten} 
\author[b]{Brian P.\ Tighe} 
\author[a]{M.\ Lisa Manning}

\affil[a]{Department of Physics, Syracuse University, Syracuse, New York 13244, USA}
\affil[b]{Delft University of Technology, Process \& Energy Laboratory, Leeghwaterstraat 39, 2628 CB Delft, The Netherlands}

\leadauthor{Merkel} 

\significancestatement{What do a guitar string and a balloon have in common?  They are both floppy unless rigidified by geometrically induced prestresses.  The same kind of rigidity transition in under-constrained materials has more recently been discussed in the context of disordered biopolymer networks and models for biological tissues.  Here, we propose a general approach to quantitatively describe such transitions.  Based on a minimal length function, which scales linearly with intrinsic fluctuations in the system and quadratically with shear strain, we make concrete predictions about the elastic response of these materials, which we verify numerically and which are consistent with previous experiments.  Finally, our approach may help develop methods that connect macroscopic elastic properties of disordered materials to their microscopic structure.}

\authorcontributions{M.M., B.P.T., and M.L.M.\ designed the research, M.M.\ performed the research and analyzed the data, K.B.\ provided important simulation data, M.M., B.P.T., and M.L.M.\ wrote the paper.}
\authordeclaration{The authors declare no conflict of interest.}
\correspondingauthor{\textsuperscript{1}To whom correspondence should be addressed. E-mail: mmerkel@syr.edu}


\begin{abstract}
We present a novel approach to understand geometric-incompatibility-induced rigidity in under-constrained materials, including sub-isostatic 2D spring networks and 2D and 3D vertex models for dense biological tissues. We show that in all these models a geometric criterion, represented by a minimal length $\bar\ell_\mathrm{min}$, determines the onset of prestresses and rigidity. This allows us to predict not only the correct scalings for the elastic material properties, but also the precise {\em magnitudes} for bulk modulus and shear modulus discontinuities at the rigidity transition as well as the magnitude of the Poynting effect. We also predict from first principles that the ratio of the excess shear modulus to the shear stress should be inversely proportional to the critical strain with a prefactor of three, and propose that this factor of three is a general hallmark of geometrically induced rigidity in under-constrained materials and could be used to distinguish this effect from nonlinear mechanics of single components in experiments. Lastly, our results may lay important foundations for ways to estimate $\bar\ell_\mathrm{min}$ from measurements of local geometric structure, and thus help develop methods to characterize large-scale mechanical properties from imaging data.
\end{abstract}

\dates{This manuscript was compiled on \today}
\doi{\url{www.pnas.org/cgi/doi/10.1073/pnas.XXXXXXXXXX}}

\begin{document}

\maketitle
\thispagestyle{firststyle}
\ifthenelse{\boolean{shortarticle}}{\ifthenelse{\boolean{singlecolumn}}{\abscontentformatted}{\abscontent}}{}

\dropcap{A} material's rigidity is intimately related to its geometry. In materials that crystallize, rigidity occurs when the constituent parts organize on a lattice.  In contrast, granular systems can rigidify while remaining disordered, and arguments developed by Maxwell \cite{Maxwell1864} accurately predict that the material rigidifies at an isostatic point where the number of constraints on particle motion equal the number of degrees of freedom. 

Further work by Calladine~\cite{Calladine1978} highlighted the important role of states of self stress, demonstrating that an index theorem relates rigidity to the total number of constraints, degrees of freedom, and self stresses.  Recent work has extended these ideas in both ordered and disordered systems to design materials with geometries that permit topologically protected floppy modes~\cite{Lubensky2015,Zhou2018,Mao2018}.

A third way to create rigidity is through geometric incompatibility, which we illustrate by a guitar string. Before it is tightened, the floppy string is under-constrained, with fewer constraints than degrees of freedom, and there are many ways to deform the string at no energetic cost. As the distance between the two ends is increased above the rest length of the string, this geometric incompatibility together with the accompanying creation of a self-stress rigidifies the system \cite{Alexander1998,Lubensky2015}. Any deformation will be associated with an energetic cost, leading to finite vibrational frequencies. 
This same mechanism has been proposed to be important for the elasticity of rubbers and gels \cite{Alexander1998} as well as biological cells \cite{Ingber2014}.  

In particular, it has been shown to rigidify under-constrained, disordered fiber networks under applied strain, with applications in biopolymer networks~\cite{Onck2005,Wyart2008,Sheinman2012,Silverberg2014a, Licup2015,Licup2016,Sharma2016a,Sharma2016b,Feng2016,Oosten2016,Vermeulen2017,Shivers2017,Shivers2018,Jansen2018,Rens2018}. Just as with the guitar string, rigidity arises when the size and shape of the box introduce external constraints that are incompatible with the local segments of the network attaining their desired rest lengths.  
For example, when applying external shear, fiber networks strongly rigidify at some critical shear strain $\gamma^\ast$ \cite{Rammensee2007,Wyart2008,Feng2016,Sharma2016a,Vermeulen2017,Shivers2017,Shivers2018,Rens2018}, although it remains controversial whether the onset of rigidity is continuous \cite{Sharma2016a,Sharma2016b,Rens2016,Shivers2018} or discontinuous \cite{Vermeulen2017} in the limit without fiber bending rigidity.  Similarly, fiber networks can also be rigidified by isotropic dilation \cite{Sheinman2012}, and the interaction between isotropic and shear elasticity in these systems is characterized an anomalous negative Poynting effect \cite{Janmey2007,DeCagny2016,Shivers2017,Baumgarten2018,Jansen2018}, i.e.\ the development of a tensile normal stress in response to externally applied simple shear.  However, it has as yet remained unclear how all of these observations and their critical scaling behavior \cite{Wyart2008,Broedersz2011,Feng2016,Vermeulen2017,Shivers2018} are quantitatively connected to the underlying geometric structure of the network. 
Moreover, while previous works have remarked that several features of stiffening in fiber networks are surprisingly independent of model details \cite{Licup2016}, it has remained elusive whether there are generic underlying mechanisms.

Rigidity transitions have also been identified in dense biological tissues~\cite{Angelini2011,Sadati2013,Park2015,Garcia2015,Malinverno2017}. In particular, vertex or Voronoi models that describe tissues as a tessellation of space into polygons or polyhedra exhibit rigidity transitions~\cite{Farhadifar2007,Staple2010,Bi2014,Bi2015,Bi2016,Matoz-Fernandez2016,Barton2017,Yang2017,Moshe2017,Giavazzi2018,Sussman2018,Sussman2018b,Merkel2018,Boromand2018,Yan2018,Teomy2018}, which share similarities with both particle-based models, where the transition is driven by changes to connectivity~\cite{Yan2018}, and fiber (or spring) networks, which can be rigidified by strain.  Therefore, an open question is how both connectivity and strain can interact to rigidify materials \cite{Rens2018}. 

\begin{table*}
  \caption{Models discussed in this article.  For the spring networks, the values indicated apply to a system size of $2N/z=1024$ nodes, and for all cellular models values apply to a system size of $N=512$ cells.
  For each model, we indicate the respective dimension $d$ of the ``length springs'' and the spatial dimension $D$, as well as the numbers of degrees of freedom (dof) as well as constraints (i.e.\ length + area springs).
  The provided values for transition point $\ell_0^\ast$ and geometric coefficients $a_\ell$, $a_a$, and $b$ are average values extracted from simulations exploring the rigid regime near the transition point. For the cellular models, they are indicated together with their standard deviations across different random realizations.  For the 2D spring networks, the indicated numbers and their uncertainty corresponds to the respective fit of the average values with fixed exponent of $\Delta z$.
  Differences to earlier publications \cite{Bi2015,Sussman2018,Merkel2018} result from differences in sampling due to a different energy minimization protocol used here (Supplemental Information, section IV).
  \label{tab:models}}
  \centering
  \setlength{\tabcolsep}{5pt}
  \renewcommand\cellgape{\Gape[1pt]}
  \begin{tabular}{lc|cc|cc|cccc}\hline
    \textbf{Model} & \textbf{``Area''} & \multicolumn{2}{c|}{\textbf{Dimension}} & \multicolumn{2}{c|}{\textbf{Number of}} & \textbf{Transition} & \multicolumn{3}{c}{\textbf{Coefficients}}
    \bigstrut[t]\\
     & \textbf{rigidity} & $d$ & $D$ & \textbf{dof} & \textbf{constraints} & \textbf{point} $\bm{\ell_0^\ast}$ & $\bm{a_\ell}$ & $\bm{a_a}$ & $\bm{b}$
     \\\hline
    2D spring network  & --  & 1 & 2 & $4N/z$ & $N$
        & \makecell{$(1.506\pm0.004)$ \\ $-(0.378\pm0.009)\Delta z$} & $(1.33\pm0.06)/\Delta z^{1/2}$ & -- & $(0.7\pm0.1)/\Delta z$ \bigstrut[t]\\\hline
    2D vertex model  & $k_A=0$ & 1 & 2 & $4N$ & $N$ 
        & $3.87\pm0.01$ & $0.30\pm0.01$ & -- & $0.48\pm0.02$ \bigstrut[t]\\
    2D vertex model  & $k_A>0$ & 1 & 2 & $4N$ & $2N$ 
        & $3.92\pm0.01$  & $1.7\pm0.4$ & $3.3\pm0.7$ & $0.6\pm0.2$ \bigstrut[t]\\\hline
    2D Voronoi model  & $k_A=0$ & 1 & 2 & $2N$ & $N$
        & $3.82\pm0.01$ & $0.64\pm0.03$ & -- & $0.68\pm0.03$ \bigstrut[t]\\\hline
    3D Voronoi model  & $k_V=0$ & 2 & 3 & $3N$ & $N$
        & $5.375\pm0.003$ & $0.25\pm0.01$ & -- & $0.61\pm0.02$ \bigstrut[t]\\
    3D Voronoi model  & $k_V>0$ & 2 & 3 & $3N$ & $2N$
        & $5.406\pm0.004$ & $2.0\pm0.1$  & $6.6\pm0.4$ & $1.1\pm0.1$ \bigstrut[t]\\\hline
  \end{tabular}
\end{table*}

Very recently, some of us showed that the 3D Voronoi model exhibits a rigidity transition driven by geometric incompatibility~\cite{Merkel2018}, similar to fiber networks.  This has also been demonstrated for the 2D vertex model, using a continuum elasticity approach based on a local reference metric \cite{Moshe2017}. For the case of the 3D Voronoi model, we found that there was a special relationship between properties of the network geometry and the location of the rigidity transition, largely independent of the realization of the disorder \cite{Merkel2018}.

\begin{figure*}
  \centering
  \includegraphics{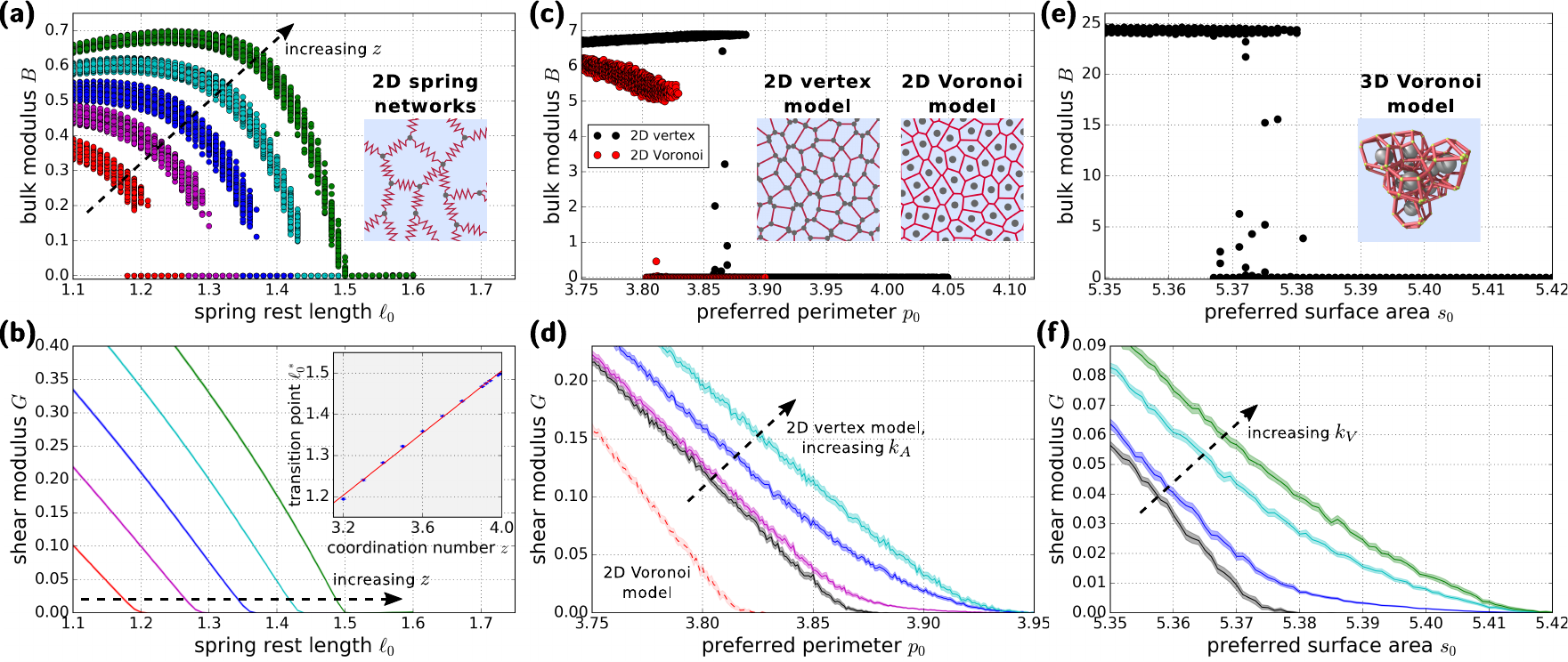}
  \caption{
  Comparison of the rigidity transition across the different models:
  (a,b) 2D spring network (coordination numbers $z=3.2,3.4,3.6,3.8,3.99$), (c,d) 2D Voronoi model (with $k_A=0$) and 2D vertex model (with $k_A=0$ in panel c and $k_A=0,0.1,1,10$ in panel d), (e,f) 3D Voronoi model (with $k_V=0$ in panel e and $k_V=0,1,10,100$ in panel f).
  In all models, the transition is discontinuous in the bulk modulus (panels a,c,e) and continuous in the shear modulus (panels b,d,f).
  (b inset)  For 2D spring networks, the value of the transition point $\ell_0^\ast$ (quantified using the bisection protocol detailed in  section IVB of the Supplemental Information) increases with the coordination number $z$. This relation is approximately linear in the vicinity of the isostatic point $z_c\equiv4$. Blue dots are simulation data and the red line shows a linear fit with $\ell_0^\ast=(1.506\pm0.004) - (0.378\pm0.009)\Delta z$ with $\Delta z = z_c-z$.
  Close to the transition point in panels c,e, data points are scattered between zero and a maximal value.  This scattering is due to insufficient energy minimization in these cases. 
  In panels b, d, and f, shaded regions indicate the standard error of the mean.
  \label{fig:shearAndBulkModuli-linear}}
\end{figure*}

Here, we show that such a relationship between rigidity and geometric structure is generic to a broad class of under-constrained materials, including spring networks and vertex/Voronoi models in different dimensions (\tref{tab:models}, \fref{fig:shearAndBulkModuli-linear}).  
We first demonstrate that all these models display the same generic behavior in response to isotropic dilation.  
Understanding key geometric structural properties of these systems allows us to predict the precise values of a discontinuity in the bulk modulus at the transition point.
We then extend our approach to include shear deformations, which allows us to analytically predict a discontinuity in the shear modulus at the onset of rigidity. Moreover, we can make precise quantitative predictions of the values of critical shear strain $\gamma^\ast$, scaling behavior of the shear modulus beyond $\gamma^\ast$, Poynting effect, and several related critical exponents.  In each case, we numerically demonstrate the validity of our approach for the case of spring networks. 

We also compare our predictions to previously published experimental data, and highlight some new predictions, including a prefactor of three that we expect to find generically in a scaling collapse of the shear modulus, shear stress, and critical strain.

We achieve these results by connecting macroscopic mechanical network properties to underlying geometric properties. In the case of the guitar string, the string first becomes taut when the distance between the two ends attains a critical value $\ell_0^{*}$ equal to the intrinsic length of the string, so that the boundary conditions for the string are geometrically incompatible with the intrinsic geometry of the string.  As the string is stretched, one can predict its pitch (or equivalently the effective elastic modulus) by quantifying the actual length of the string $\ell$ relative to its intrinsic length.  While this is straightforward in the one-dimensional geometry of a string, we are interested in understanding whether a similar geometric principle, based on the average length of a spring $\bar\ell$ governs the behavior near the onset of rigidity in disordered networks in 2D and 3D.

Here, we formulate a geometric compatibility criterion in terms of the constrained minimization of the average spring length $\bar\ell_\mathrm{min}$  in a disordered network. Just as for the guitar string, this length $\bar\ell_\mathrm{min}$ attains a critical value $\ell_0^{*}$ at the onset of rigidity. As the system is strained beyond the rigidity transition, we demonstrate analytically and numerically that the geometry constrains $\bar\ell_\mathrm{min}$ to vary in a simple way with two observables: fluctuations of spring lengths $\sigma_l$, and shear strain $\gamma$. Because $\bar\ell_\mathrm{min}$ is minimized over the whole network, it is a collective geometric property of the network. 

Just as with the guitar string, the description of the geometry given by $\bar\ell_\mathrm{min}$ then allows us to calculate many features of the elastic response, including the bulk and shear moduli. This in turn provides a general basis to analytically understand the strain-stiffening responses of under-constrained materials to both isotropic and anisotropic deformation within a common framework.
Even though $\bar\ell_\mathrm{min}$ describes collective geometric effects, our work may also provide an important foundation to understand macroscopic mechanical properties from \emph{local} geometric structure.

\section{Models}
Here we focus on four classes of models, which include 2D sub-isostatic random spring networks without bending rigidity \cite{Wyart2008,Tighe2012,Yucht2013,During2013,During2014,Woodhouse2018} and three models for biological tissues: the 2D vertex model \cite{Farhadifar2007,Bi2015}, the 2D Voronoi model \cite{Bi2016,Sussman2018}, and the 3D Voronoi model \cite{Merkel2018} (\tref{tab:models}). 

2D spring networks consist of nodes that are connected by in total $N$ springs, where the average number of springs connected to a node is the coordination number $z$. We create networks with a defined value for $z$ by translating jammed configurations of bidisperse disks into spring networks and then randomly pruning springs until the desired coordination number $z$ is reached \cite{Wyart2008,Baumgarten2018}.  We use harmonic springs, such that the total mechanical energy of the system is:
\begin{equation}
  e_{s2D} = \sum_i{(l_i-l_{0i})^2}\text{.}\label{eq:energy-spring-network-1}
\end{equation}
Here, the sum is over all springs $i$ with length $l_i$ and rest length $l_{0i}$, which are generally different for different springs. 
For convenience, we re-express \eref{eq:energy-spring-network-1} in terms of a mean spring rest length $\ell_0=[(\sum_i{l_{0i}^2})/N]^{1/2}$, which we use as a control parameter acting as a common scaling factor for all spring rest lengths.  This allows us to rewrite the energy as:
\begin{equation}
  e_{s2D} = \sum_i{w_i(\ell_i-\ell_{0})^2}\label{eq:energy-spring-network-2}
\end{equation}
with rescaled spring lengths $\ell_i=\ell_0l_i/l_{0i}$ and weights $w_i=(l_{0i}/\ell_0)^2$, such that $\sum_i{w_i}=N$ (for details, see Supplemental Information, section IA).
In simple constraint counting arguments, each spring is treated as one constraint, and here we are interested in \emph{sub-isostatic} (i.e\ under-constrained, also called \emph{hypostatic}) networks with $z<z_c\equiv4$.

The tissue models describe biological tissues as polygonal (2D) or polyhedral (3D) tilings of space.  For the Voronoi models, these tilings are Voronoi tessellations and the degrees of freedom are the Voronoi centers of the cells.  In contrast, in the 2D vertex model, the degrees of freedom are the positions of the vertices (i.e.\ the polygon corners).
Forces between the cells are described by an effective energy functional. For the 2D models, the (dimensionless) energy functional is:
\begin{equation}
  e_{c2D} = \sum_{i}{\biggl[(p_i-p_0)^2 + k_A(a_i-1)^2\biggr]} \text{.}\label{eq:energy2D}
\end{equation}
Here, the sum is over all $N$ cells $i$ with perimeter $p_i$ and area $a_i$.  There are two parameters in this model: the preferred perimeter $p_0$ and the relative area elasticity $k_A$. 
For the 3D Voronoi model, the energy is defined analogously:
\begin{equation}
  e_{c3D} = \sum_{i}{\biggl[(s_i-s_0)^2 + k_V(v_i-1)^2\biggr]}\text{.}\label{eq:energy3D}
\end{equation}
The sum is again over all $N$ cells $i$ of the configuration, with cell surface area $s_i$ and volume $v_i$, and the two parameters of the model are preferred surface area $s_0$ and relative volume elasticity $k_V$. 

All four of these models are under-constrained based on simple constraint counting, as is apparent from the respective numbers of degrees of freedom and constraints listed in \tref{tab:models}.
We stress that Calladine's constraint counting derivation \cite{Calladine1978,Lubensky2015} also applies to many-particle, non-central-force interactions. 

Throughout this article, we will often discuss all four models at once.  Thus, when generally talking about ``elements'', we refer to springs in the spring networks and cells in the tissue models. Similarly, when talking about ``lengths $\ell$'' (of dimension $d$), we refer to spring lengths $\ell$ in the spring networks, cell perimeters $p$ in the 2D tissue models, and cell surface areas $s$ in the 3D tissue model (\tref{tab:models}).  
Finally, when talking about ``areas $a$'' (of dimension $D$), we refer to cell areas $a$ in the 2D tissue models as well as cell volumes $v$ in the 3D tissue model. 

Here we study the behavior of local energy minima of all four models under periodic boundary conditions with fixed dimensionless system size $N$, i.e.\ the model is non-dimensionalized such that the average area per element is one \cite{Yang2017,Merkel2018,Sussman2018}.
Under these conditions, a rigidity transition exists in all models even without area rigidity.  In particular, for the 2D vertex and 3D Voronoi models, we discuss the special case $k_A=0$ separately (\tref{tab:models}).
Moreover, the athermal 2D Voronoi model does not exhibit a rigidity transition for $k_A>0$ \cite{Sussman2018}, and thus we will only discuss the case $k_A=0$ for this model. 

\section{Results}
\subsection{Rigidity is created by geometric incompatibility corresponding to a minimal length criterion}
We start by comparing the rigidity transitions in the four different models using \fref{fig:shearAndBulkModuli-linear}, where we plot both the differential bulk modulus $B$ and the differential shear modulus $G$ versus the preferred length $\ell_0$.
In this first part, we use for all models the preferred length $\ell_0$ as a control parameter. Note that because $\ell_0$ is non-dimensionalized using the number density of elements, changing $\ell_0$ corresponds to applying isotropic strain (i.e.~a change in volume with no accompanying change in shape).  Later, we will additionally include the shear strain $\gamma$ as a control parameter.

In all models, we find a rigid regime ($B,G>0$) for preferred lengths below the transition point $\ell_0^\ast$, and a floppy regime ($B=G=0$) above it, with the transition being discontinuous in the bulk modulus and continuous in the shear modulus.
For the spring networks, we find that the transition point $\ell_0^\ast$ depends on the coordination number, where close to the isostatic point $z_c\equiv4$, it scales linearly with the distance $\Delta z=z_c-z$ to isostaticity (\fref{fig:shearAndBulkModuli-linear}b inset), as previously similarly discussed in \cite{Sheinman2012}. Something similar has also been reported for a 2D vertex model \cite{Yan2018}.

For the cellular models, we find that the transition point for the case without area rigidity, $k_A=0$, is generally smaller than in the case with area rigidity, $k_A>0$ (\fref{fig:shearAndBulkModuli-linear}d,f, \tref{tab:models}).
Moreover, our 2D vertex model transition point for $k_A>0$ is somewhat higher than reported before \cite{Bi2015}.  Here we used a different vertex model implementation than in \cite{Bi2015} (Supplemental Information, section IVC), and the location of the transition in vertex models depends somewhat on the energy minimization protocol \cite{Sussman2018}, a feature that is shared with other models for disordered materials \cite{Chaudhuri2010}.
Also, in \fref{fig:shearAndBulkModuli-linear}d,f the \emph{averaged} shear modulus always becomes zero at a higher value than the respective average transition point listed in \tref{tab:models}.  This is due to the distribution of transition points having a finite width (see also finite width of $\ell_0$ regions with both zero and nonzero bulk moduli in panels c and e).

We find that in all these models, the mechanism creating the transition is the same: rigidity is created by geometric incompatibility, which is indicated by the existence of prestresses.  We have already shown this for the 3D Voronoi model \cite{Merkel2018} and the 2D Voronoi model with $k_A=0$ \cite{Sussman2018}, while others have shown this for the \emph{ordered} 2D vertex model~\cite{Moshe2017}. Furthermore, our data confirms that this is the case for the 2D spring networks and the $k_A=0$ cases of both (disordered) 2D vertex and 3D Voronoi models (Supplemental Information, section IIA).  

We find something similar for the disordered 2D vertex model for $k_A>0$.  Although there are special cases where prestresses appear also in the floppy regime (Supplemental Information, section IIA), to simplify our discussion here, we only consider configurations without such typically localized prestresses.

We observe that in all of these models, a geometric criterion, which we describe in terms of a minimal average length $\bar\ell_\mathrm{min}$, determines the onset of prestresses. For example, we can exactly transform the spring network energy \eref{eq:energy-spring-network-2} into (Supplemental Information, section IA):
\begin{equation}
  e_{s2D} = N\Bigl[(\bar\ell-\ell_0)^2+\sigma_\ell^2 \Bigr]\text{.}\label{eq:energy-simplified}
\end{equation}
Here, $\bar\ell=(\sum_i{w_i\ell_i})/N$ and $\sigma_\ell^2=(\sum_i{w_i(\ell_i-\bar\ell)^2})/N$ are weighted average and standard deviation of the \emph{rescaled} spring lengths.
This means that $\bar\ell$ and $\sigma_\ell$ are average and standard deviation of the actual spring lengths $l_i$\emph{, each measured relative to its actual rest length $l_{0i}$}.
In particular, the standard deviation $\sigma_\ell$ vanishes whenever all springs $i$ have the same value of the \emph{fraction} $l_i/l_{0i}$, even though the absolute lengths $l_i$ may differ among the springs.
Moreover, importantly, the mean rest length $\ell_0$ enters the definitions of $\bar\ell$ and $\sigma_\ell$, but only via the ratios $l_{0i}/\ell_0$, which characterize the \emph{relative} spring length distribution. Hence, the ``rescaled'' geometric information contained in both $\bar\ell$ and $\sigma_\ell$ is a combination of the actual spring lengths and the \emph{relative} rest length distribution, but is independent of the \emph{absolute} mean rest length $\ell_0$.

According to \eref{eq:energy-simplified}, energy minimization corresponds to a simultaneous minimization with respect to $\lvert \bar\ell-\ell_0\rvert$ and $\sigma_\ell$:
In the floppy regime we find numerically that both quantities can vanish simultaneously and thus, all lengths attain their rest lengths, $\ell_i=\ell_0$ (Supplemental Information, section IIA).  
In contrast in the rigid regime, $\lvert \bar\ell-\ell_0\rvert$ and $\sigma_\ell$ cannot both simultaneously vanish, creating tensions $2(\ell_i-\ell_0)$, which are sufficient to rigidify the network.
The transition point $\ell_0^\ast$ corresponds to the smallest possible preferred spring length $\ell_0$ for which the system can still be floppy.  In other words, it corresponds to a local minimum in the average rescaled spring length $\ell_0^\ast=\mathrm{min}\,{\bar\ell}$ of the network under the constraint of no fluctuations of the rescaled lengths, $\sigma_\ell=0$.  Because this minimization is with respect to all node positions and includes all springs, it defines the distribution of transition points $\ell_0^\ast$ as a collective property of the rescaled geometry of 2D spring networks.

For the cellular models with $k_A>0$, we analogously find that the transition point is given by the minimal cell perimeter $\bar\ell$ (surface in 3D) under the constraint of no cell perimeter \emph{and area fluctuations} $\sigma_\ell=\sigma_a=0$, which now additionally appear in the energy \eref{eq:energy-simplified} \cite{Merkel2018}.
Again, this is a geometric criterion, which also explains why the transition point $\ell_0^\ast$ is independent of $k_A$ for $k_A>0$ (\fref{fig:shearAndBulkModuli-linear}d,f).
Moreover, we can understand why the transition point is smaller for $k_A=0$: in this case the energy does not constrain the area fluctuations, and the transition point is given by the minimal perimeter under the weaker constraint of having no perimeter fluctuations.  Thus, the transition point will generally be smaller for the $k_A=0$ case than for the $k_A>0$ case. 

\begin{figure*}
  \centering
  \includegraphics{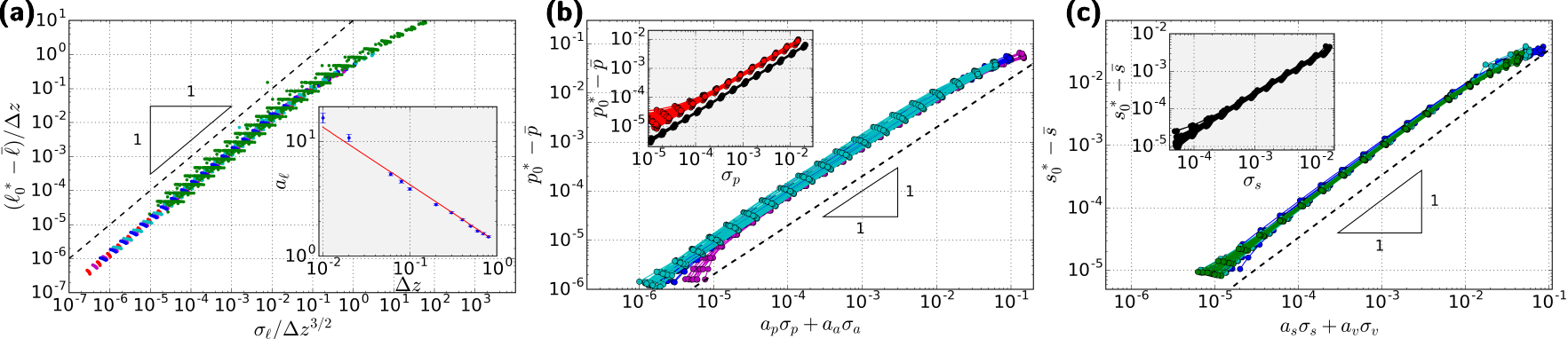}
  \caption{Verification of the geometric linearity near the transition point. The difference between average length and transition point, $\ell_0^\ast-\bar{\ell}$, scales linearly with the standard deviations of lengths $\sigma_\ell$ and areas $\sigma_a$.
  (a) 2D spring network, (b) 2D Voronoi and vertex models, (c) 3D Voronoi model.  The values of $z$, $k_A$, and $k_V$ are respectively as in \fref{fig:shearAndBulkModuli-linear}.
  (a inset) For the 2D spring networks, the coefficient $a_\ell$ in \eref{eq:length-scaling-kA=0} scales with the distance to isostaticity approximately as $a_\ell\sim\Delta z^{-1/2}$.
  In all panels, deviations from linearity exist for large $\ell_0^\ast-\bar{\ell}$ because \esref{eq:length-scaling-kA=0}~and~\seref{eq:length-scaling} describe the behavior close to the transition point, and deviations for small $\ell_0^\ast-\bar{\ell}$ are due the finite cutoff on the shear modulus used to obtain the transition point value $\ell_0^\ast$ (Supplemental Information, section IV).
  \label{fig:geometry-scaling}}
\end{figure*}
\begin{figure*}
  \centering
  \includegraphics{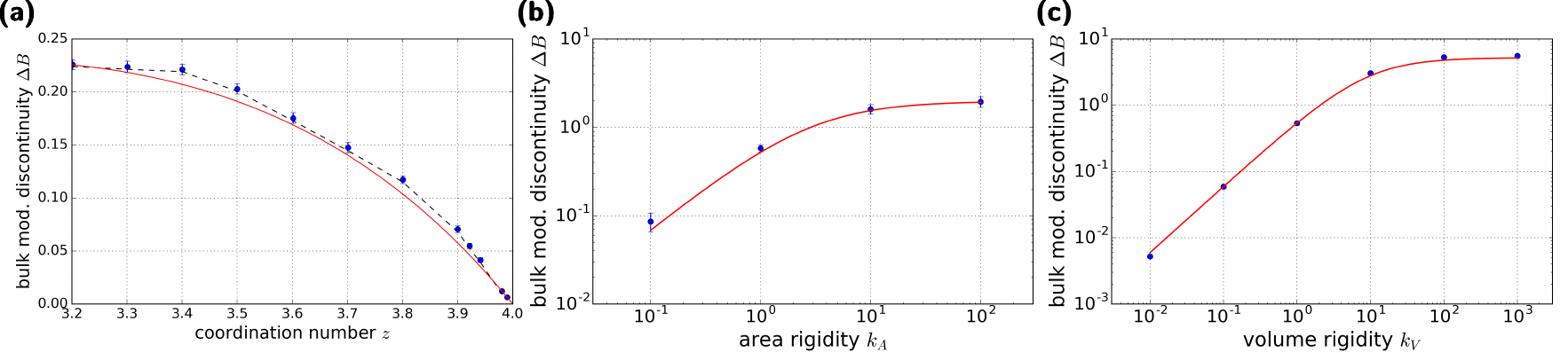}
  \caption{Predicted and observed behavior of the bulk modulus discontinuity $\Delta B$ for (a) 2D spring networks for different values of the coordination number $z$, (b) the 2D vertex model for different values of the area rigidity $k_A$ and (c) the 3D Voronoi model for different values of the volume rigidity $k_V$.
  Blue dots indicate simulations and the red curves indicate predictions without fit parameters based on \eref{eq:bulkModulusPrediction}.  In panel a, the black dashed curve is computed using values for transition point $\ell_0^\ast$ and geometric scaling coefficient $a_\ell$ directly measured for each value of $z$, while for the red line we used the scaling relations from \tref{tab:models}.
  \label{fig:bulkModulus-kA}}
\end{figure*}
\subsection{The minimal length scales linearly with fluctuations}
We next study the scaling of the minimal length in the rigid vicinity of the transition.  In the rigid regime, the system must compromise between minimizing $\lvert \bar\ell-\ell_0\rvert$ and $\sigma_\ell$ (and possibly $\sigma_a$ in cellular models). To understand how, we must account for geometric constraints, which we express in terms of how the minimal length $\bar\ell_\mathrm{min}=\mathrm{min}\,{\bar\ell}$ depends on the fluctuations: $\bar\ell_\mathrm{min}=\bar\ell_\mathrm{min}(\sigma_\ell,\sigma_a)$. In the rigid regime the observed average length is always greater than the preferred length, $\bar\ell>\ell_0$, and so the average length instead takes on its locally minimal possible value $\bar\ell=\bar\ell_\mathrm{min}(\sigma_\ell,\sigma_a)$. Therefore, knowing the functional form of $\bar\ell_\mathrm{min} (\sigma_\ell,\sigma_a)$ will allow us to predict how the system energy $e$ (and thus also the bulk and shear moduli) depend on the control parameter $\ell_0$ (Supplemental Information, section IC-E).

In section IB of the supplement, we show analytically that in the absence of prestresses in the floppy regime, the minimal length $\bar\ell_\mathrm{min}$ depends linearly on the standard deviations 
$\sigma_\ell$ and $\sigma_a$.  This is directly related to the state of self-stress that is created at the onset of geometric incompatibility at $\ell_0=\ell_0^\ast\equiv\bar\ell_\mathrm{min}(0,0)$ \cite{Lubensky2015}. 

To check this prediction, we numerically simulate these models, and observe indeed a linear scaling of the $\bar\ell_\mathrm{min}(\sigma_\ell)$ functions close to the transition point (\fref{fig:geometry-scaling}).  In particular, for 2D spring networks and the $k_A=0$ cases of the cellular models, we find:
\begin{equation}
  \bar\ell_\mathrm{min}(\sigma_\ell) = \ell_0^\ast - a_\ell\sigma_\ell \label{eq:length-scaling-kA=0}
\end{equation}
with scaling coefficient $a_\ell$.  We list its value in \tref{tab:models} for the different models.
Interestingly, we find that the coefficient $a_\ell$ is largely independent of the random realization of the system, in particular for cellular models with $k_A=0$.

For 2D spring networks, $a_\ell$ depends on the coordination number $z$ and approximately scales as $a_\ell\sim\Delta z^{-1/2}$ (\fref{fig:geometry-scaling}a inset).  This scaling behavior of $a_\ell$ can be rationalized using a scaling argument based on the density of states (Supplemental Information, section IF).

For cellular models where area plays a role, \eref{eq:length-scaling-kA=0} is extended (\fref{fig:geometry-scaling}b,c):
\begin{equation}
  \bar\ell_\mathrm{min}(\sigma_\ell,\sigma_a) = \ell_0^\ast - a_\ell\sigma_\ell - a_a\sigma_a\text{.}\label{eq:length-scaling}
\end{equation}
Again the coefficients $a_\ell$ and $a_a$ are listed in \tref{tab:models} for 2D vertex and 3D Voronoi models.
The coefficients $a_\ell$ differ significantly between the $k_A>0$ and $k_A=0$ cases of the same model, which makes sense because \esref{eq:length-scaling-kA=0} and \seref{eq:length-scaling} are linear expansions of the function $\bar\ell_\mathrm{min}(\sigma_\ell,\sigma_a)$ at different points $(\sigma_\ell,\sigma_a)$.  

\subsection{Prediction of the bulk modulus discontinuity}
Knowing the behavior of the minimal length function $\bar\ell_\mathrm{min}(\sigma_\ell,\sigma_a)$ in the rigid phase near the transition point provides us with an explicit expression for the energy in terms of the control parameter $\ell_0$ (Supplemental Information, section IC):
\begin{equation}
  e(\ell_0) = \frac{N}{Z}(\ell_0^\ast-\ell_0)^2
\end{equation}
with $Z=1+a_\ell^2+a_a^2/k_A$, where for models without an area term the $a_a^2/k_A$ term is dropped.
Because changes in $\ell_0$ correspond to changes in system size, we can predict the exact value of the bulk modulus discontinuity, $\Delta B$, at the transition in all models (\fref{fig:shearAndBulkModuli-linear}a-c, Supplemental Information, section IE):
\begin{equation}
  \Delta B = \frac{2d^2(\ell_0^\ast)^2}{D^2Z}\text{.}\label{eq:bulkModulusPrediction}
\end{equation}
This equation is for a model with $d$-dimensional ``lengths'' embedded in a $D$-dimensional space (see \tref{tab:models}). For the special case of a hexagonal lattice in the 2D vertex model, this result is consistent with Ref.~\cite{Murisic2015}. More generally, for disordered networks the geometric coefficients $a_\ell$ and $a_a$ appear in the denominator, because they describe non-affinities that occur in response to global isotropic deformations (Supplemental Information, section IE).
A comparison of the predicted $\Delta B$ to simulation results is shown in \fref{fig:bulkModulus-kA}.

\subsection{Nonlinear elastic behavior under shear}
\begin{figure*}
  \centering
  \includegraphics{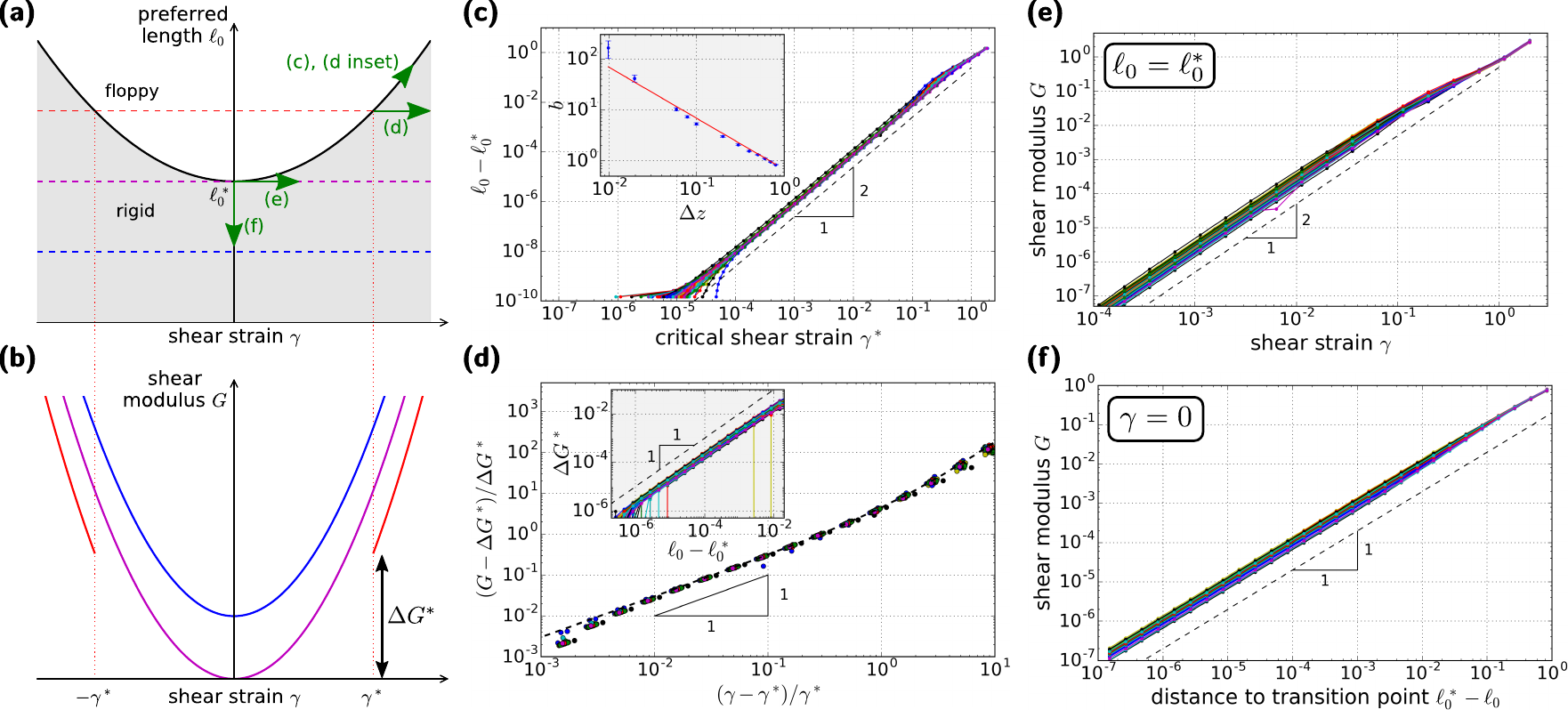}
  \caption{Nonlinear elastic behavior of sub-isostatic spring networks under shear. 
  (a) Schematic phase diagram illustrating the parabolic boundary between rigid (shaded) and floppy (unshaded) regime depending on preferred spring length $\ell_0$ and shear strain $\gamma$.
  (b) Schematic showing the dependence of the shear modulus $G$ on the shear strain $\gamma$ for different values of $\ell_0$ (cf.\ panel a).  Note that for $\ell_0>\ell_0^\ast$ (red curve), \eref{eq:G} predicts a discontinuity $\Delta G^\ast$ in the shear modulus at the onset of rigidity. 
  (c) We numerically find a quadratic dependence between $\ell_0-\ell_0^\ast$ and the critical shear $\gamma^\ast$ where the network rigidifies for given $\ell_0>\ell_0^\ast$.  This is consistent with our Taylor expansion in \eref{eq:lmin-with-gamma}, and the quadratic regime extends to shear strains of up to $\gamma\sim0.1$. 
  Deviations for very small $\ell_0-\ell_0^\ast$ are attributed to the finite shear modulus cutoff of $10^{-10}$ used to probe the phase boundary (Supplemental Information, section IVB).
  (c inset) The prefactor $b$ associated with the quadratic relation in panel c scales approximately as $b\sim1/\Delta z$.  
  (d) Scaling of the shear modulus beyond the shear modulus discontinuity, $(G-\Delta G^\ast)/\Delta G^\ast$ over $(\gamma-\gamma^\ast)/\gamma^\ast$ with $\ell_0-\ell_0^\ast=10^{-4}$.  The dashed black line indicates the prediction from \eref{eq:G} without fit parameters.
  (d inset) Scaling of the shear modulus discontinuity $\Delta G^\ast$ with $\ell_0-\ell_0^\ast$.
  (e,f) Scaling of the shear modulus with $\gamma$ and $\ell_0^\ast-\ell_0$, respectively.  In all panels the coordination number is $z=3.2$.
  \label{fig:parabola}}
\end{figure*}
As shown before \cite{Onck2005,Wyart2008,Sheinman2012, Licup2015,Sharma2016a,Sharma2016b,Feng2016,Vermeulen2017,Shivers2017,Shivers2018,Jansen2018}, under-constrained systems can also be rigidified by applying finite shear strain.  We now incorporate shear strain $\gamma$ into our formalism and test our predictions on the 2D spring networks.  However, we expect our findings to equally apply to the cell-based models (Supplemental Information, section IC,D).
We also numerically verified that our analytical predictions also apply to 2D fiber networks without bending rigidity (Supplemental Information, section IIC).

To extend our approach, we take into account that the minimal-length function $\bar\ell_\mathrm{min}(\sigma_\ell)$ can in principle also depend on the shear strain $\gamma$. We thus Taylor expand in $\gamma$:
\begin{equation}
  \bar\ell_\mathrm{min}(\sigma_\ell,\gamma) = \ell_0^\ast - a_\ell\sigma_\ell + b\gamma^2\text{,}\label{eq:lmin-with-gamma}
\end{equation}
where the linear term in $\gamma$ is dropped due to symmetry when expanding about an isotropic state (in practice, for our finite-sized systems we drop the linear term in $\gamma$ by defining the $\gamma=0$ point using shear stabilization, Supplemental Information, sections ID and IV).
While at the moment we have no formal proof that $\ell_\mathrm{min}$ is analytic, and the ultimate justification for \eref{eq:lmin-with-gamma} comes from a numerical check (see next paragraph), we hypothesize that for most systems $\ell_\mathrm{min}$ will be analytic in $\gamma$, up to randomly scattered points $\gamma$ where singularities in the form of plastic rearrangements occur.

For a fixed value of $\gamma$, the interface between solid and rigid regime is again given by $\bar\ell_\mathrm{min}(\sigma_\ell=0,\gamma)$, and the corresponding phase diagram in terms of both control parameters $\gamma$ and $\ell_0$ is illustrated in \fref{fig:parabola}a.  Indeed, we also numerically find a quadratic scaling for the transition line, $\ell_0-\ell_0^\ast=b(\gamma^\ast)^2$, extending up to shear strains of $\gamma\sim0.1$ (\fref{fig:parabola}c, see also Supplemental Information, section IIB).
We find that for spring networks the coefficient $b$ depends on $\Delta z$ approximately as $b\sim\Delta z^{-1}$ (\fref{fig:parabola}c inset), which can be understood from properties of the density of states (Supplemental Information, section IF).
To optimize precision, values of $b$ have been extracted from the relation $G=4b(\bar\ell-\ell_0)$ in this plot (see below, cf.\ \fref{fig:parabola}f).

Knowing the functional form of $\bar\ell_\mathrm{min}(\sigma_\ell,\gamma)$ close to the transition line allows us to explicitly express the energy in the rigid regime in terms of both control parameters (Supplemental Information, section IC):
\begin{equation}
  e(\ell_0,\gamma) = \frac{N}{1+a_\ell^2}\Big(\ell_0^\ast-\ell_0+b\gamma^2\Big)^2\text{.}\label{eq:energy-l0-gamma}
\end{equation}
This allows us to explicitly compute the shear modulus $G=(\mathrm{d}^2e/\mathrm{d}\gamma^2)/N$. We obtain for both floppy and rigid regime:
\begin{equation}
  G(\ell_0,\gamma) = \Theta\Big(\ell_0^\ast-\ell_0+b\gamma^2\Big)\frac{4b}{1+a_\ell^2}\Big(\ell_0^\ast - \ell_0 + 3b\gamma^2 \Big)\text{,}\label{eq:G}
\end{equation}
where $\Theta$ is the Heaviside function.
We now discuss several consequences of this expression for the shear modulus (\fref{fig:parabola}b).

When shearing the system starting in the floppy regime (i.e.\ for $\ell_0>\ell_0^\ast$), \eref{eq:G} predicts a \emph{discontinuous} change in the shear modulus of $\Delta G^\ast=8b(\ell_0-\ell_0^\ast)/(1+a_\ell^2)$ at the onset of rigidity at $\gamma^\ast=[(\ell_0-\ell_0^\ast)/b]^{1/2}$.  We verify the linear scaling $\Delta G^\ast\sim(\ell_0-\ell_0^\ast)$ in \fref{fig:parabola}d inset, and the value of the scaling coefficient in the Supplemental Information, section IIB.  Moreover, \eref{eq:G} also correctly predicts the behavior beyond $\gamma^\ast$, as shown in  \fref{fig:parabola}d.  

\eref{eq:G} also correctly predicts the shear modulus behavior for $\ell_0\leq\ell_0^\ast$.  For $\ell_0=\ell_0^\ast$, the shear modulus scales quadratically with $\gamma$ (\fref{fig:parabola}e), while for $\gamma=0$, the shear modulus scales linearly with $(\ell_0^\ast-\ell_0)>0$ (\fref{fig:parabola}f, see Supplemental Information, section ID, for the cellular models), as reported before for many of the cellular models \cite{Bi2015,Murisic2015,Merkel2018}.  In both cases, we verified that the respective coefficients coincide with their expected values based on the values of $a_\ell$ and $b$.

In particular for $\gamma=0$, because $(\ell_0^\ast-\ell_0)=(1+a_\ell^2)(\bar\ell-\ell_0)$, we obtain the simple relation $G=4b(\bar\ell-\ell_0)$, which explains the collapse in the shear modulus scaling for different $k_V$ in the 3D Voronoi model that some of us reported earlier \cite{Merkel2018}.

\begin{figure}
  \centering
  \includegraphics{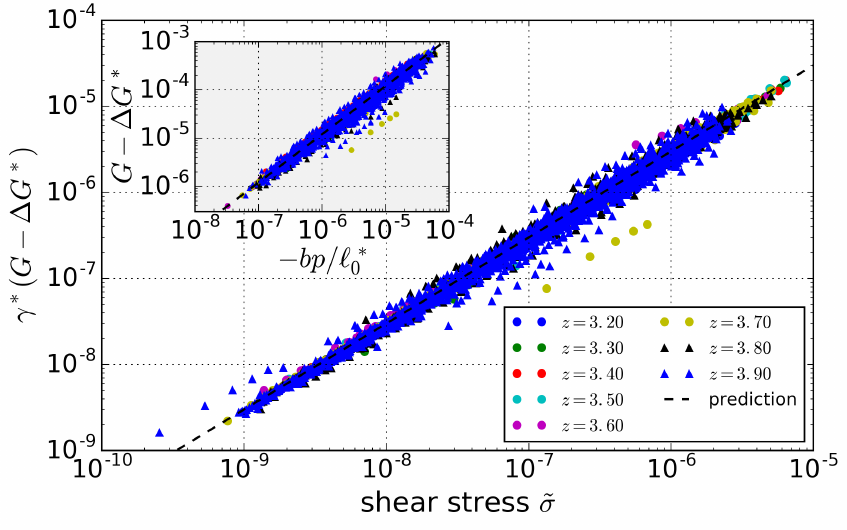}
  \caption{The excess shear modulus $G-\Delta G^\ast$ scales linearly with the shear stress $\tilde\sigma$ in 2D spring networks.  We find a collapse when rescaling $G-\Delta G^\ast$ by the critical shear strain $\gamma^\ast$.  The black dashed line corresponds to the prefactor of 3, as predicted by \eref{eq:G-stresses}.  (inset) The excess shear modulus $G-\Delta G^\ast$ scales linearly with the isotropic stress $-p$, and we obtain a collapse when rescaling the latter by $b/\ell_0^\ast$.  The black dashed line is the prediction according to \eref{eq:G-stresses}.  
  \label{fig:shearModulus-stress}}
\end{figure}
We also obtain explicit expressions for both shear stress $\tilde\sigma=(\mathrm{d}e/\mathrm{d}\gamma)/N$ and isotropic stress, i.e.\ negative pressure $-p$ (Supplemental Information, sections ID,E).  For the latter, we find a negative Poynting effect with coefficient $\chi \equiv p/\gamma^2= -2db\ell_0^\ast/D(1+a_\ell^2)$ at $\ell_0=\ell_0^\ast$.  Moreover, we find the following relations for the shear modulus:
\begin{align}
  G &= \Delta G^\ast + \frac{3}{\gamma}\tilde\sigma &
  G &= \Delta G^\ast - \frac{6Db}{d\ell_0^\ast}p\text{.}\label{eq:G-stresses}
\end{align}
Indeed, we observe a collapse of our simulation data for the 2D spring networks in both cases (\fref{fig:shearModulus-stress} \& inset), where we use that close to the onset of rigidity, $\gamma\simeq\gamma^\ast$.

\section{Discussion}
In this article, we propose a unifying perspective on under-constrained materials that are stiffened by geometric incompatibility.  This is relevant for a broad class of materials \cite{Alexander1998}, and has more recently been discussed in the context of biopolymer gels \cite{Onck2005,Licup2015,Licup2016,Sharma2016a,Jansen2018} and biological tissues \cite{Bi2015,Park2015,Merkel2018,Moshe2017}. Just as with a guitar string, we are able to predict many features of the mechanical response of these systems by quantifying geometric incompatibility -- we develop a generic geometric rule $\bar\ell_\mathrm{min}$ for how generalized springs in a disordered network deviate from their rest length. Using this minimal average length function $\bar\ell_\mathrm{min}$, we then derive the macroscopic elastic properties of a very broad class of under-constrained, prestress-rigidified materials from first principles.  
We numerically verify our findings using models for biopolymer networks \cite{Wyart2008,Sharma2016a} and biological tissues \cite{Farhadifar2007,Bi2016,Merkel2018}.

Our work is relevant for experimentalists and may explain the reproducibility of a number of generic mechanical features found in particular for biopolymer networks \cite{Janmey2007,Licup2015,Oosten2016,Jansen2018}.
While we neglect here a fiber bending rigidity that is included in many biopolymer network models \cite{Licup2015,Licup2016,Sharma2016a,Sharma2016b,Jansen2018}, future work that includes such a term will further refine our theoretical results and the following comparison to experiments (see below).
For shear deformations with $\ell_0$ sufficiently close to $\ell_0^\ast$ and close to the onset of rigidity $\gamma\simeq\gamma^\ast$, we predict a linear scaling of the differential shear modulus $G$ with the shear stress $\tilde{\sigma}$, where $(G-\Delta G^\ast)/\tilde{\sigma}\sim1/\gamma^\ast$, which has been reported before for biopolymer networks \cite{Licup2015,Licup2016,Jansen2018}.  However, here we additionally predict from first principles that the value of the prefactor is exactly 3, a factor consistent with previous experimental results \cite{Licup2015,Jansen2018}. Moreover, our work strongly suggests that the relation $(G-\Delta G^\ast)/\tilde{\sigma}=3/\gamma$ is a general hallmark of prestress-induced rigidity in under-constrained materials.  We thus propose it as a general experimental criterion to test whether an observed strain-stiffening behavior can be understood in terms of geometrically induced rigidity.  If applicable to biopolymer gels, this could help to discern whether strain-stiffening of a gel is due to the nonlinear mechanics of single filaments or is dominated by prestresses, a long-standing question in the field \cite{Storm2005,Onck2005}.

We can also apply these predictions to typical rheometer geometries (Supplemental Information, section IG).
We predict that an atypical tensile normal stress $\sigma_{zz}$ develops under simple shear, which corresponds to a negative Poynting effect, that $\sigma_{zz}$ scales linearly with shear stress and shear modulus: $\sigma_{zz}\sim\tilde\sigma\sim(G-\Delta G^\ast)$ (\eref{eq:G-stresses} and Supplemental Information, section IG). 
This is precisely what has been found for many biopolymer gels like collagen, fibrin, or matrigel \cite{Janmey2007,Licup2015,DeCagny2016,Jansen2018}. 
However, in contrast to Ref.~\cite{Jansen2018}, our work suggests that the scaling factor between $\sigma_{zz}$ and $(G-\Delta G^\ast)$ should be largely independent of $\gamma^\ast$.
While these effects can also be explained by nonlinearities \cite{Storm2005,Janmey2007,Kang2009,Cioroianu2013}, and have already been discussed in the context of prestress-induced rigidity \cite{Licup2016,Shivers2017,Jansen2018}, we show here that they represent a very generic feature of prestress-induced rigidity in under-constrained materials.

Our work also highlights the importance of isotropic deformations when studying prestress-induced rigidity, as demonstrated experimentally in Ref.~\cite{Oosten2016}.  While previous work \cite{Onck2005,Wyart2008,Licup2015,Sharma2016a,Sharma2016b,Vermeulen2017,Shivers2018,Jansen2018} focused almost \cite{Sheinman2012} entirely on shear deformations, we additionally study the effect of isotropic deformations represented by the control parameter $\ell_0$.  First, due to the bulk modulus discontinuity, our work predicts zero normal stress under compression and linearly increasing normal stress under expansion, consistent with experimental findings on biopolymer networks \cite{Oosten2016} (assuming the uniaxial response is dominated by the isotropic part of the stress tensor, see Supplemental Information, section IG). 
Second, we also correctly predict that the critical shear strain $\gamma^\ast$ increases upon compression, which corresponds to an increase in $\ell_0$ \cite{Oosten2016} (cf.\ \fref{fig:parabola}a).
While we also predict an increase of the shear modulus $G$ under extension, which was observed as well \cite{Oosten2016}, additional effects arising from the superposition of pure shear and simple shear very likely play an important role in this case. While we consider this outside the scope of this article, it will be straight-forward to extend our work by this aspect.

In summary, we have developed a new approach to understand how many under-constrained disordered materials rigidify in a manner similar to a guitar string. 
While it is clear that the one-dimensional string becomes rigid precisely when it is stretched past its rest length, we show that in two- and three-dimensional models, rigidity is governed by a geometrical minimal length function $\bar\ell_\mathrm{min}$ with generic features (e.g.\ linear scaling with intrinsic fluctuations, quadratic scaling with shear strain).  This insight allows us to make accurate predictions for many of the scaling functions and prefactors that describe the linear response of these materials.  In addition, by performing numerical measurements of the geometry in the rigid phase to extract the coefficients of the $\bar\ell_\mathrm{min}$ function, we can even predict the precise magnitudes of several macroscopic mechanical properties.

In addition, these predictions help unify or clarify several scaling collapses that have been identified previously in the literature. For 2D spring networks derived from jammed packings, we studied the dependence of our geometric coefficients on the coordination number $z$, and find that approximately, $a_\ell\sim\Delta z^{-1/2}$ and $b\sim\Delta z ^{-1}$.  
Combined with our finding that the value of $\ell_0$ right after initialization depends linearly on $z$, such that $(\ell_0-\ell_0^\ast)\sim\Delta z$ (Figure~S5a~inset in the Supplemental Information), we obtain that the critical shear strain $\gamma^\ast$ scales as $\gamma^\ast\sim\Delta z^\beta$ with $\beta=1$.  Similarly, we find for the associated shear modulus discontinuity $\Delta G^\ast\sim\Delta z^\theta$ with $\theta=1$.  While both exponents are consistent with earlier findings by Wyart et al.~\cite{Wyart2008}, our approach highlights the importance of the initial value of $\ell_0$ for the elastic properties under shear.
In other work, bond-diluted regular networks yielded different exponents $\beta$ and $\theta$ \cite{Feng2016}, which is not surprising because the scaling exponents of $a_\ell$ and $b$ with $\Delta z$ are likely dependent on the way the network is generated.
More generally, while we observed that the values of $\ell_0^\ast$, $a_\ell$, $a_a$, and $b$ depended somewhat on the protocol of system preparation and energy minimization, they were relatively reproducible among different random realizations of a given protocol \cite{Chaudhuri2010}.

Moreover, we analytically predict and numerically confirm 
the existence and precise value of a shear modulus discontinuity $\Delta G^\ast$ with respect to shear deformation, whose existence for fiber networks without bending rigidity has been controversially discussed more recently \cite{Sharma2016a,Sharma2016b,Rens2016,Vermeulen2017,Shivers2018}.  We also predict a generic scaling of the shear modulus beyond this discontinuity: $(G-\Delta G^\ast)\sim(\gamma-\gamma^\ast)^f$ with $f=1$. 
Smaller values for $f$ that have been reported before for different kinds of spring and fiber networks \cite{Sharma2016a,Sharma2016b,Vermeulen2017,Shivers2018} are likely due to higher order terms in $\bar\ell_\mathrm{min}$.
Given the very generic nature of our approach, we expect to find a value of $f=1$ in these systems as well, if probed sufficiently close to $\ell_0=\ell_0^\ast$. 

One major obstacle in determining elastic properties of disordered materials is the appearance of non-affinities, which can lead to a break-down of approaches like effective medium theory close to the transition \cite{Sheinman2012}.  
In our case, effects by non-affinities are by construction fully included in the geometric coefficients $a_\ell$, $a_a$, and $b$.
However, while measures for non-affinity have been discussed before \cite{Wyart2008,Broedersz2011,Broedersz2012,Sharma2016b,Shivers2018}, these are usually quite distinct from our coefficients $a_\ell$, $a_a$, and $b$.  For example for spring networks, such earlier definitions typically include spring \emph{rotations}, while our coefficients represent changes in spring \emph{length} only. 
Hence, while earlier definitions reflect much of the actual \emph{motion} of the microscopic elements, our coefficients only retain the part directly relevant for the system energy and thus the mechanics.  
In other words, the coefficients $a_\ell$, $a_a$, and $b$ (and $\ell_0^\ast$) can be regarded as a minimal set of parameters required to characterize the elastic system properties close to the transition.

There are a number of possible future extensions of this work.  
First, we have focused here on transitions created by a minimal length, where the system is floppy for large $\ell_0$ and rigid for small $\ell_0$.  However, there is in principle also the possibility of a transition created by e.g.\ a maximal length, which is for example the case in classical sphere jamming.  Although we have occasionally seen something like this in our spring networks close to isostaticity, we generally expect this to be less typical in under-constrained systems due to buckling.

Second, while we studied here the vicinity of one local minimum of $\bar\ell_\mathrm{min}$ depending e.g.\ on $\gamma$, it would be interesting to study the behavior of the system beyond that, by including higher order terms in $\bar\ell_\mathrm{min}$, and by also explicitly taking plastic events into account \cite{Amuasi2018}.  In the case of biological tissues, plastic events typically correspond to so-called T1 transitions \cite{Kim2018}, which in our approach would correspond to changing to a different $\bar\ell_\mathrm{min}$ ``branch''.

Third, it will be important to study what determines the exact values of the geometric coefficients $a_\ell$, $a_a$, and $b$, how they depend on the network statistics, and why they are relatively reproducible.  For the cellular models with area term, preliminary results suggest that the ratio of both ``$a$'' coefficients can be estimated by $a_a/a_\ell\approx d\ell_0^\ast/D$, because the self-stress that appears at the onset of rigidity seems to be dominated by a force balance between cell perimeter tension and pressure within each cell. 

Fourth, because we separated geometry from energetics, it is in principle possible to generalize our work to other interaction potentials, e.g.\ the correct expression for semi-flexible filaments \cite{Storm2005,Cioroianu2013}, and to include the effect of active stresses \cite{Ronceray2015,Stam2017,Woodhouse2018,Fischer-Friedrich2018}.  Note that our work directly generalizes to any analytic interaction potential with a local minimum at a finite length.  Although in this more general case \eref{eq:energy-simplified} would include higher order cumulants of $\ell_i$, these higher order terms will be irrelevant in the floppy regime and we expect them to be negligible in the rigid vicinity of the transition, where we make most of our predictions.

Fifth, this work may also provide foundations to systematically connect macroscopic mechanical material properties to the underlying \emph{local} geometric structure.  
For example for biopolymer networks, properties of the local geometric structure can be extracted using light scattering, scanning electron microscopy, or confocal reflectance microscopy \cite{Roeder2002,Lindstrom2010,Jansen2018}.
In particular, our simulations indicate that in models without area term the $\bar\ell_\mathrm{min}$ function does not change much when increasing system size by nearly an order of magnitude (Supplemental Information, section IID), which suggests that local geometry may indeed be sufficient to characterize the large-scale mechanical properties of such systems.
Remaining future challenges here include the development of an easy way to compute our geometric coefficients from simple properties characterizing local geometric structure without the need to simulate, and to find ways to detect possible residual stresses that may have been built into the gel during polymerization.

Finally, our approach can likely be extended to also include isostatic and over-constrained materials.
For example, it is generally assumed that the mechanics of biopolymer networks is dominated by a stretching rigidity of fibers that form a sub-isostatic network, but that an additional fiber bending rigidity turns the network into an over-constrained system \cite{Licup2015,Licup2016,Sharma2016a,Sharma2016b,Jansen2018,Rens2018}.  The predictions we make here focus on the stretching-dominated limit where fiber bending rigidity can be neglected, which is attained by a weak fiber bending modulus and/or in the more rigid parts of the phase space.
A generalization of our formalism towards over-constrained systems will allow us to extend our predictions beyond this regime and thus refine our comparison to experimental data.

\acknow{We thank Daniel M.\ Sussman for fruitful discussions. MM and MLM acknowledge funding from the Simons Foundation under grant number 446222, the Alfred P.\ Sloan Foundation, the Gordon and Betty Moore Foundation, the Research Corporation for Scientific Advancement though the Cottrell Scholars program, and computational support through NSF ACI-1541396. MLM also acknowledges support from the Simons Foundation under grant number 454947, and NSF-DMR-1352184 and NSF-PHY-1607416. KB and BPT acknowledge funding from the Netherlands Organization for Scientific Research (NWO).}

\matmethods{\subsection*{Numerical implementation of the models}
The 2D spring networks were initialized as packing-derived, randomly cut networks \cite{Wyart2008,Baumgarten2018}.
To improve the precision as compared to the cellular models, we created our own implementation of the Polak-Ribi\`ere version of the conjugate gradient minimization method \cite{Nash1995}, where for the line searches we use a self-developed Newton method based only on energy derivatives.  All states were minimized until the average force per degree of freedom was less than $10^{-12}$.  For the $\ell_0$ sweep in \fref{fig:shearAndBulkModuli-linear}a,b and to find the $(\gamma,\ell_0)=(0,\ell_0^\ast)$ point, we used  shear stabilization.  Details are given in section IVB of the Supplemental Information.

For the 2D vertex model simulations, we always started from Voronoi tessellations of random point patterns, generated using the Computational Geometry Algorithms Library (CGAL, \url{https://www.cgal.org/}), and we used the BFGS2 implementation of the GNU Scientific Library (GSL,  \url{https://gnu.org/software/gsl/}) to minimize the energy.  We enforced 3-way vertices and the length cutoff for T1 transitions was set to $10^{-5}$, and there is a maximum possible number of T1 transitions on a single cell-cell interface of $10^4$. 
All 2D vertex model configurations studied were shear stabilized.

For the 2D Voronoi model simulations, we started from random point patterns and minimized the system energy using the BFGS2 routine of the GSL, each time using CGAL to compute the Voronoi tessellations.  Due to limitations of CGAL, configurations were not shear stabilized.

For the 3D Voronoi model simulations, we used the shear-stabilized, energy-minimized states generated in Ref.~\cite{Merkel2018} using the BFGS2 multidimensional minimization routine of the GSL.

Details on the different simulation protocols ($\ell_0$ sweeps and bisection to obtain the transition point) are discussed in detail in section IV of the Supplemental Information.}



\showacknow{} 

\bibliography{references}

\begin{thebibliography}{10}

\bibitem{Maxwell1864}
Maxwell JC (1864) {On the calculation of the equilibrium and stiffness of
  frames}.
\newblock {\em Philosophical Magazine Series 4} 27(182):294--299.

\bibitem{Calladine1978}
Calladine CR (1978) {Buckminster Fuller's "Tensegrity" structures and Clerk
  Maxwell's rules for the construction of stiff frames}.
\newblock {\em International Journal of Solids and Structures} 14(2):161--172.

\bibitem{Lubensky2015}
Lubensky TC, Kane CL, Mao X, Souslov A, Sun K (2015) {Phonons and elasticity in
  critically coordinated lattices}.
\newblock {\em Reports on Progress in Physics} 78(7):73901.

\bibitem{Zhou2018}
Zhou D, Zhang L, Mao X (2018) {Topological Edge Floppy Modes in Disordered
  Fiber Networks}.
\newblock {\em Physical Review Letters} 120(6):68003.

\bibitem{Mao2018}
Mao X, Lubensky TC (2018) {Maxwell Lattices and Topological Mechanics}.
\newblock {\em Annu. Rev. Condens. Matter Phys} 9:413--33.

\bibitem{Alexander1998}
Alexander S (1998) {Amorphous solids: Their structure, lattice dynamics and
  elasticity}.
\newblock {\em Physics Report} 296(2-4):65--236.

\bibitem{Ingber2014}
Ingber DE, Wang N, Stamenovi{\'{c}} D (2014) {Tensegrity, cellular biophysics,
  and the mechanics of living systems}.
\newblock {\em Reports on Progress in Physics} 77(4):046603.

\bibitem{Onck2005}
Onck PR, Koeman T, van Dillen T, van~der Giessen E (2005) {Alternative
  Explanation of Stiffening in Cross-Linked Semiflexible Networks}.
\newblock {\em Physical Review Letters} 95(17):178102.

\bibitem{Wyart2008}
Wyart M, Liang H, Kabla A, Mahadevan L (2008) {Elasticity of floppy and stiff
  random networks}.
\newblock {\em Physical Review Letters} 101(21):1--4.

\bibitem{Sheinman2012}
Sheinman M, Broedersz CP, MacKintosh FC (2012) {Nonlinear effective-medium
  theory of disordered spring networks}.
\newblock {\em Physical Review E} 85(2):021801.

\bibitem{Silverberg2014a}
Silverberg JL, et~al. (2014) {Structure-Function Relations and Rigidity
  Percolation in the Shear Properties of Articular Cartilage}.
\newblock {\em Biophysical Journal} 107(7):1721--1730.

\bibitem{Licup2015}
Licup AJ, et~al. (2015) {Stress controls the mechanics of collagen networks}.
\newblock {\em Proceedings of the National Academy of Sciences}
  112(31):9573--9578.

\bibitem{Licup2016}
Licup AJ, Sharma A, MacKintosh FC (2016) {Elastic regimes of subisostatic
  athermal fiber networks}.
\newblock {\em Physical Review E} 93(1):012407.

\bibitem{Sharma2016a}
Sharma A, et~al. (2016) {Strain-controlled criticality governs the nonlinear
  mechanics of fibre networks}.
\newblock {\em Nature Physics} 12(6):584--587.

\bibitem{Sharma2016b}
Sharma A, et~al. (2016) {Strain-driven criticality underlies nonlinear
  mechanics of fibrous networks}.
\newblock {\em Physical Review E} 94(4):042407.

\bibitem{Feng2016}
Feng J, Levine H, Mao X, Sander LM (2016) {Nonlinear elasticity of disordered
  fiber networks}.
\newblock {\em Soft Matter} 12(5):1419--1424.

\bibitem{Oosten2016}
van Oosten ASG, et~al. (2016) {Uncoupling shear and uniaxial elastic moduli of
  semiflexible biopolymer networks: compression-softening and
  stretch-stiffening}.
\newblock {\em Scientific Reports} 6(1):19270.

\bibitem{Vermeulen2017}
Vermeulen MFJ, Bose A, Storm C, Ellenbroek WG (2017) {Geometry and the onset of
  rigidity in a disordered network}.
\newblock {\em Physical Review E} 96(5):053003.

\bibitem{Shivers2017}
Shivers J, Feng J, Sharma A, MacKintosh FC (2017) {Anomalous normal stress
  controlled by marginal stability in fiber networks}.
\newblock {\em arXiv:1711.00522}.

\bibitem{Shivers2018}
Shivers J, Arzash S, Sharma A, MacKintosh FC (2018) {Scaling theory for
  mechanical critical behavior in fiber networks}.
\newblock {\em arXiv:1807.01205} (c).

\bibitem{Jansen2018}
Jansen KA, et~al. (2018) {The Role of Network Architecture in Collagen
  Mechanics}.
\newblock {\em Biophysical Journal} 114(11):2665--2678.

\bibitem{Rens2018}
Rens R, Villarroel C, D{\"{u}}ring G, Lerner E (2018) {Micromechanical theory
  of strain-stiffening of biopolymer networks}.
\newblock {\em arXiv:1808.04756}.

\bibitem{Rammensee2007}
Rammensee S, Janmey PA, Bausch AR (2007) {Mechanical and structural properties
  of in vitro neurofilament hydrogels}.
\newblock {\em European Biophysics Journal} 36(6):661--668.

\bibitem{Rens2016}
Rens R, Vahabi M, Licup AJ, MacKintosh FC, Sharma A (2016) {Nonlinear Mechanics
  of Athermal Branched Biopolymer Networks}.
\newblock {\em The Journal of Physical Chemistry B} 120(26):5831--5841.

\bibitem{Janmey2007}
Janmey PA, et~al. (2007) {Negative normal stress in semiflexible biopolymer
  gels}.
\newblock {\em Nature Materials} 6(1):48--51.

\bibitem{DeCagny2016}
de~Cagny HCG, et~al. (2016) {Porosity Governs Normal Stresses in Polymer Gels}.
\newblock {\em Physical Review Letters} 117(21):217802.

\bibitem{Baumgarten2018}
Baumgarten K, Tighe BP (2018) {Normal Stresses, Contraction, and Stiffening in
  Sheared Elastic Networks}.
\newblock {\em Physical Review Letters} 120(14):148004.

\bibitem{Broedersz2011}
Broedersz CP, Mao X, Lubensky TC, MacKintosh FC (2011) {Criticality and
  isostaticity in fibre networks}.
\newblock {\em Nature Physics} 7(12):983--988.

\bibitem{Angelini2011}
Angelini TE, et~al. (2011) {Glass-like dynamics of collective cell migration}.
\newblock {\em Proceedings of the National Academy of Sciences}
  108(12):4714--4719.

\bibitem{Sadati2013}
Sadati M, {Taheri Qazvini} N, Krishnan R, Park CY, Fredberg JJ (2013)
  {Collective migration and cell jamming}.
\newblock {\em Differentiation} 86(3):121--125.

\bibitem{Park2015}
Park JA, et~al. (2015) {Unjamming and cell shape in the asthmatic airway
  epithelium}.
\newblock {\em Nature Materials} 14(10):1040--1048.

\bibitem{Garcia2015}
Garcia S, et~al. (2015) {Physics of active jamming during collective cellular
  motion in a monolayer}.
\newblock {\em Proceedings of the National Academy of Sciences}
  112(50):15314--15319.

\bibitem{Malinverno2017}
Malinverno C, et~al. (2017) {Endocytic reawakening of motility in jammed
  epithelia}.
\newblock {\em Nature Materials} 16(5):587--596.

\bibitem{Farhadifar2007}
Farhadifar R, R{\"{o}}per JC, Aigouy B, Eaton S, J{\"{u}}licher F (2007) {The
  Influence of Cell Mechanics, Cell-Cell Interactions, and Proliferation on
  Epithelial Packing}.
\newblock {\em Current Biology} 17(24):2095--2104.

\bibitem{Staple2010}
Staple DB, et~al. (2010) {Mechanics and remodelling of cell packings in
  epithelia}.
\newblock {\em The European Physical Journal E} 33(2):117--127.

\bibitem{Bi2014}
Bi D, Lopez JH, Schwarz JM, Manning ML (2014) {Energy barriers and cell
  migration in densely packed tissues}.
\newblock {\em Soft Matter} 10(12):1885.

\bibitem{Bi2015}
Bi D, Lopez JH, Schwarz JM, Manning ML (2015) {A density-independent rigidity
  transition in biological tissues}.
\newblock {\em Nature Physics} 11(12):1074--1079.

\bibitem{Bi2016}
Bi D, Yang X, Marchetti MC, Manning ML (2016) {Motility-Driven Glass and
  Jamming Transitions in Biological Tissues}.
\newblock {\em Physical Review X} 6(2):021011.

\bibitem{Matoz-Fernandez2016}
Matoz-Fernandez DA, Martens K, Sknepnek R, Barrat JL, Henkes S (2017) {Cell
  division and death inhibit glassy behaviour of confluent tissues}.
\newblock {\em Soft Matter} 13(17):3205--3212.

\bibitem{Barton2017}
Barton DL, Henkes S, Weijer CJ, Sknepnek R (2017) {Active Vertex Model for
  cell-resolution description of epithelial tissue mechanics}.
\newblock {\em PLOS Computational Biology} 13(6):e1005569.

\bibitem{Yang2017}
Yang X, et~al. (2017) {Correlating Cell Shape and Cellular Stress in Motile
  Confluent Tissues}.
\newblock {\em Proceedings of the National Academy of Sciences}
  114(48):12663--12668.

\bibitem{Moshe2017}
Moshe M, Bowick MJ, Marchetti MC (2017) {Geometric frustration and solid-solid
  transitions in model 2D tissue}.
\newblock {\em Physical Review Letters} 120(26):268105.

\bibitem{Giavazzi2018}
Giavazzi F, et~al. (2018) {Flocking Transitions in Confluent Tissues}.
\newblock {\em Soft Matter}.

\bibitem{Sussman2018}
Sussman DM, Merkel M (2018) {No unjamming transition in a Voronoi model of
  biological tissue}.
\newblock {\em Soft Matter}.

\bibitem{Sussman2018b}
Sussman DM, Paoluzzi M, {Cristina Marchetti} M, {Lisa Manning} M (2018)
  {Anomalous glassy dynamics in simple models of dense biological tissue}.
\newblock {\em EPL (Europhysics Letters)} 121(3):36001.

\bibitem{Merkel2018}
Merkel M, Manning ML (2018) {A geometrically controlled rigidity transition in
  a model for confluent 3D tissues}.
\newblock {\em New Journal of Physics} 20(2):022002.

\bibitem{Boromand2018}
Boromand A, Signoriello A, Ye F, O'Hern CS, Shattuk M (2018) {Jamming of
  Deformable Polygons}.
\newblock {\em arXiv:1801.06150}.

\bibitem{Yan2018}
Yan L, Bi D (2018) {Rosette-driven rigidity transition in epithelial tissues}.
\newblock {\em arXiv:1806.04388}.

\bibitem{Teomy2018}
Teomy E, Kessler DA, Levine H (2018) {Confluent and non-confluent phases in a
  model of cell tissue}.
\newblock {\em arXiv:1803.03962} (2).

\bibitem{Tighe2012}
Tighe BP (2012) Dynamic critical response in damped random spring networks.
\newblock {\em Physical Review Letters} 109(16):168303.

\bibitem{Yucht2013}
Yucht M, Sheinman M, Broedersz C (2013) Dynamical behavior of disordered spring
  networks.
\newblock {\em Soft Matter} 9(29):7000--7006.

\bibitem{During2013}
D{\"{u}}ring G, Lerner E, Wyart M (2013) {Phonon gap and localization lengths
  in floppy materials}.
\newblock {\em Soft Matter} 9(1):146--154.

\bibitem{During2014}
D{\"u}ring G, Lerner E, Wyart M (2014) Length scales and self-organization in
  dense suspension flows.
\newblock {\em Physical Review E} 89(2):022305.

\bibitem{Woodhouse2018}
Woodhouse FG, Ronellenfitsch H, Dunkel J (2018) Autonomous actuation of zero
  modes in mechanical networks far from equilibrium.
\newblock {\em arXiv:1805.07728}.

\bibitem{Chaudhuri2010}
Chaudhuri P, Berthier L, Sastry S (2010) {Jamming transitions in amorphous
  packings of frictionless spheres occur over a continuous range of volume
  fractions}.
\newblock {\em Physical Review Letters} 104(16):165701.

\bibitem{Murisic2015}
Murisic N, Hakim V, Kevrekidis IG, Shvartsman SY, Audoly B (2015) {From
  Discrete to Continuum Models of Three-Dimensional Deformations in Epithelial
  Sheets}.
\newblock {\em Biophysical Journal} 109(1):154--163.

\bibitem{Storm2005}
Storm C, Pastore JJ, MacKintosh FC, Lubensky TC, Janmey PA (2005) {Nonlinear
  elasticity in biological gels}.
\newblock {\em Nature} 435(7039):191--194.

\bibitem{Kang2009}
Kang H, et~al. (2009) {Nonlinear Elasticity of Stiff Filament Networks: Strain
  Stiffening, Negative Normal Stress, and Filament Alignment in Fibrin Gels
  †}.
\newblock {\em The Journal of Physical Chemistry B} 113(12):3799--3805.

\bibitem{Cioroianu2013}
Cioroianu AR, Storm C (2013) {Normal stresses in elastic networks}.
\newblock {\em Physical Review E} 88(5):052601.

\bibitem{Broedersz2012}
Broedersz CP, Sheinman M, MacKintosh FC (2012) {Filament-length-controlled
  elasticity in 3D fiber networks}.
\newblock {\em Physical Review Letters} 108(7):3--7.

\bibitem{Amuasi2018}
Amuasi H, Fischer A, Zippelius A, Heussinger C (2018) {Linear rheology of
  reversibly cross-linked biopolymer networks}.
\newblock {\em arXiv:1808.05407}.

\bibitem{Kim2018}
Kim S, Wang Y, Hilgenfeldt S (2018) {Universal Features of Metastable State
  Energies in Cellular Matter}.
\newblock {\em Physical Review Letters} 120(24):248001.

\bibitem{Ronceray2015}
Ronceray P, Broedersz CP, Lenz M (2016) {Fiber networks amplify active stress}.
\newblock {\em Proceedings of the National Academy of Sciences}
  113(11):2827--2832.

\bibitem{Stam2017}
Stam S, et~al. (2017) {Filament rigidity and connectivity tune the deformation
  modes of active biopolymer networks}.
\newblock {\em Proceedings of the National Academy of Sciences}
  114(47):E10037--E10045.

\bibitem{Fischer-Friedrich2018}
Fischer-Friedrich E (2018) {Active Prestress Leads to an Apparent Stiffening of
  Cells through Geometrical Effects}.
\newblock {\em Biophysical Journal} 114(2):419--424.

\bibitem{Roeder2002}
Roeder BA, Kokini K, Sturgis JE, Robinson JP, Voytik-Harbin SL (2002) {Tensile
  Mechanical Properties of Three-Dimensional Type I Collagen Extracellular
  Matrices With Varied Microstructure}.
\newblock {\em Journal of Biomechanical Engineering} 124(2):214.

\bibitem{Lindstrom2010}
Lindstr{\"{o}}m SB, Vader DA, Kulachenko A, Weitz DA (2010) {Biopolymer network
  geometries: Characterization, regeneration, and elastic properties}.
\newblock {\em Physical Review E} 82(5):051905.

\end{thebibliography}

\end{document}